# Towards More Security in Data Exchange:
# Defining Unparsers with Context-Sensitive Encoders for Context-Free Grammars

Lars Hermerschmidt, Stephan Kugelmann, Bernhard Rumpe

*Software Engineering*

*RWTH Aachen University, Germany*

{*hermerschmidt, rumpe*}@se-rwth.de, *stephan.kugelmann@rwth-aachen.de*

*http://www.se-rwth.de/*

*Abstract*—To exchange complex data structures in distributed systems, documents written in context-free languages are exchanged among communicating parties.

Unparsing these documents correctly is as important as parsing them correctly because errors during unparsing result in injection vulnerabilities such as cross-site scripting (XSS) and SQL injection. Injection attacks are not limited to the web world. Every program that uses input to produce documents in a context-free language may be vulnerable to this class of attack. Even for widely used languages such as HTML and JavaScript, there are few approaches that prevent injection attacks by context-sensitive encoding, and those approaches are tied to the language.

Therefore, the aim of this paper is to derive context-sensitive encoder from context-free grammars to provide correct unparsing of maliciously crafted input data for all context-free languages. The presented solution integrates encoder definition into context-free grammars and provides a generator for context-sensitive encoders and decoders that are used during (un)parsing. This unparsing process results in documents where the input data does neither influence the structure of the document nor change their intended semantics. By defining encoding during language definition, developers who use the language are provided with a clean interface for writing and reading documents written in that language, without the need to care about security-relevant encoding.

## I. INTRODUCTION

IT systems constantly exchange information in complex formats, which requires thoughtful design and agreement among communicating parties. In order to define these formats, it is a good idea to use context-free grammars because parsers for such languages can be constructed as classic kinds of automata recognizing these languages [1]. Moreover, such recognizer-based parsers avoid hard or undecidable problems inherent in typical ad hoc parsers. Analogously, sending information works like parsing: by unparsing the system's internal representation of information into a document [2].

Unfortunately, developers aim for easy-to-use solutions that can be implemented with low effort. Grammars, (un)parsers, and strong data types are not the first choice when building software that reads or writes documents conforming to context-free languages. The seeming simplicity of creating documents by treating all kinds of values as strings and substituting them into templates lures developers. This trend is most notable for XML-based formats even though it negates the actual structuring aspect of using XML. This simplification results in systems that work fine for most cases as long as security is not considered a critical success factor. However from a security viewpoint, the simplifications in (un)parsing introduced by developers are vulnerabilities during input data processing that can be exploited to control the behavior of the system [1] or, in case of unparsing, influence the created document in a way not intended by the developer. Well-known instances of these input based vulnerabilities are buffer overflows, SQL injections, cross-site scripting (XSS), etc. Injection attacks are not limited to the web world. Every program that uses input data to produce a document in a context-free language may be vulnerable to this class of attack.

To correctly embed input data into a document, encoders, also called sanitizers, must be used during unparsing as they escape or filter data that is used as a keyword or control token in the language, so that the developer-intended semantics of the document cannot be changed by maliciously crafted input data. As complex languages like HTML have multiple contexts, where each context requires a different encoding of input data, encoding has to be done during unparsing, because the context is unknown in advance. Developers who use unstructured string substitution or templates to assemble documents have to manually determine the context within the document and apply the correct context-specific encoder to prevent input injection. An empirical analysis by Saxena, et. al. [3] as well as countless input-based vulnerabilities classified by SANS and OWASP [4], [5] show this method of processing to be very error-prone.

Research on approaches to ease the processing of formal languages has a long history starting with generating parsers from grammars [6] and automatically formatting documents [7], [8], which lead to the more general concept of unparsing [2].

Efforts on preventing injection attacks for the languages HTML and JavaScript solved cross-site scripting more or less successfully [9] by introducing context-sensitive encoder [10]. However, these encoders have to be used for languages other than HTML as well, since input-based security vulnerabilities have been identified in cars, industrial control systems, and almost all other IT systems.

To ease the development and the correct use of context-sensitive encoders for custom languages, we aim to derive context-sensitive encoders and decoders from context-free grammars and integrate them into the (un)parsing or process to provide correct unparsing of maliciously crafted input data for all context-free languages. The approach presented in this paper integrates the encoder definition into context-free grammars and provides a generator to derive unparsers, context-sensitive encoders, and parsers as well as context-sensitive decoders. The generated language processing tools provide an interface to read and write documents, enforcing that input data neither influences the structure of the document nor change its intended semantics. Therefore, developers who use the generated tools do not need to care about security-relevant encoding and injection attacks are prevented from the beginning.

By integrating encoding into the language definition, our approach shifts away the problem of correct encoding from the application developer to the language developer who should be aware of all specialties introduced into the language. In case application developers decide to let bigger parts of the output document be controlled by input data as a whole, they need to enter the field of language development.

The remainder of this paper is structured as follows: We first review related work. In Section III, we analyze the issue of correct unparsing and encoding of maliciously crafted input data. Our approach to this problem is presented in Section IV, including composition of languages and their feature reduction to use them as input data. We evaluate this approach by taking the example of HTML and JavaScript in Section V and draw an outlook and conclusion in Section VI.

## II. RELATED WORK

Parsing [6] and unparsing [2], [7], [8] are research fields with a long history. Unparsers are also called pretty-printers if they produce pretty looking documents. Pretty-printers are generated from context-free grammars [11] and vice versa [12] to avoid redundancy and inconsistency [13].

Several language workbenches exist [14] that support developers during language development and processing. The Rascal meta-programming language [15] is used to parse, analyze, transform and unparse source code. In addition, it supports the use of a grammar of an output document to ensure syntactically correct unparsing of the document's parse tree. Hence, Rascal offers helpful tools for implementing our approach. Another language workbench, similar to the MontiCore [16] approach for defining languages that we use in this work, is the Eclipse-based Xtext [17], which uses context-free grammars for language definition and generates language processing tools for the Eclipse IDE.

To help developers implement secure systems, the Open Web Application Security Project (OWASP) aims to raise awareness [4] for web application security, provides education [18], and develops libraries for the correct processing of user input in web applications [19].

Different frameworks exist to automatically prevent XSS. Weinberger et al. [9] studied 14 of these frameworks and found only one sanitizing input data within all HTML, JavaScript, CSS, and URI contexts. Analyzing 8 popular open source web applications they found nearly all using all contexts within HTML related languages, which emphasizes the demand for context-sensitive encoders. In order to fix cross-site scripting in existing applications with only minimal modification, XSS-Guard [20] instruments the application and sends a request twice to the application: the first time with the actual user input and the second time with a safe dummy input. Subsequently, XSS detection works by comparing the responses and searching for DOM differences. Hence XSS-Guard assumes that users are not allowed to insert any HTML into the response.

A more constructive way of preventing injection attacks is to wrap around the language (e.g. SQL) a strong typed API to produce documents in this language like SQL DOM [21] or Java prepared statements do. Bravenboer et al. generalized this approach with StringBorg [22], which generates a strongly typed API for an arbitrary guest language (e.g. SQL) from its grammar in a chosen host language (e.g. Java). In addition, StringBorg embeds the syntax of the guest language into the host language, providing developers with a fluent integration of the two languages, while at the same time requiring both a complete grammar of both the guest and host languages.

Another more formal approach by Samuel et al. [10] proposes and implements a type system for template languages used in web applications. The type system is used to determine the HTML context when user input is inserted into a template and subsequently the sensitization engine calls the correct context-sensitive encoder for the input data. A similar approach from the unparsing research field by Danielsson [23] proposes an unparser which enforces correct data types in the unparsed document.

Our presented approach shares the fundamental notion with existing work on Language-theoretic security (LangSec) [24] that shotgun parsers [1], [25] will not solve the input validation problem. In addition, we perceive that input handling does not stop after parsing but is a challenge when producing output as well. Applying formal approaches, like the ones already proposed [1], [24], [26], to unparsing looks promising.

Blitzableiter [27] by Felix "FX" Lindner uses strict parsing to safely recognize the SWF format and then unparses a SWF document from the correctly parsed content, which is subsequently passed to the vulnerable Flash Player. The strict parsing of Blitzableiter validates the entire SWF structure and hence mitigates several vulnerabilities in Flash Player that could be exploited by maliciously crafted SWF content. With Blitzableiter, our approach shares the idea of validating data on the AST and consequently producing only correctly formatted data.

In summary, there are libraries and frameworks for correctly processing input data in HTML, some to be applied manually in the correct context, some for automatically detecting contexts. Furthermore, there are language workbenches for easing the development of new languages. Although generating parsers and unparsers is not a new approach, and type systems for unparsers and templates have been proposed, there is no approach for deriving context-sensitive encoders from context-free grammars.

## III. The Problem of Correct Unparsing and Encoding

To analyze the core problem behind injection attacks like SQL injection and XSS, we take a look at how documents written in a given language are used for communication in distributed systems.

First, the language used for communication should be defined by a context-free grammar [24], so that documents can be formally recognized, i.e. accepted as a member of the language or rejected. When receiving a document, a system uses a parser to read the information contained in the document into a parse tree. This tree is then refined to an abstract syntax tree (AST) that is used by system developers within the program logic to access the information from the document. Sending information analogously works by unparsing the data provided by the developer within the AST into a document. We follow the definition by Danielsson [23] and Arnoldus et al. [28] of a correct (un)parse round-trip, i.e., the (un)parsing process for a given language is a correct round-trip, if for every AST $x$ holds $parse(unparse(x)) = x$. However, we extend this property as follows:

Within the program logic, it is common to pass information from the input to another system using some language $L$. For example, data from an HTTP request is passed into a SQL statement or directly into an HTML response. An adversary who sends a specially crafted document to the system might exploit injection vulnerabilities within all components in the processing logic, resulting in an AST $m$ of the output document that contains maliciously crafted input data. The data in AST $m$ is considered malicious if $\nexists d \in L : parse(d) = m$ holds. An injection vulnerability in this context means that the adversary is able to produce documents in the language $L$ that are not intended by the developer. Parsers and unparsers that have a correct round-trip for the AST $m$ containing *maliciously crafted data* prevent this class of vulnerability by correctly encoding data from the AST into a document and safely parsing the document: $parse_{decode}(unparse_{encode}(m)) = m$. This property of a correct (un)parse round-trip explicitly assumes that data tokens within the AST may contain maliciously crafted data, whereas Danielsson and Arnoldus et al. assumed the AST to contain only valid (previously parsed) data within the data tokens.

In contrast to this formal unparsing process, developers often use more lightweight mechanisms like templates and string substitution and concatenation to produce documents, as these mechanisms do not require a formal definition of the produced document's language up front. Although this approach is easier to use, it falls short when maliciously crafted input data is inserted into a template, as the data may contain control tokens [29] of the language, which are inserted into the resulting document. Parsing this document will result in a different AST and hence exploit an injection vulnerability, as the unparse process was not correct. One commonly used solution is to implement encoder functions that remove control tokens from data or encode them using escape sequences. As different contexts within a language have different control tokens, each context needs an individual encoder. For example JavaScript requires a different encoding when used within an HTML script tag context than in an onclick-attribute context. So when using input data within a template, developers have to correctly identify the context into which the data is written and apply the corresponding encoder. If an encoder is used in a context it was not made for, it may encode some control characters by coincidence, but fail to encode others. This can represent an injection vulnerability.

The existence of countless input based vulnerabilities [4], [5] and an empirical analysis by Saxena et.al. [3] emphasize that many developers need help in solving this problem. Although considerable effort has been spent to develop encoding libraries [19] and frameworks [9] for wide-spread languages like HTML and SQL, these cannot be reused if another language is used for communication. This is a pressing issue especially in domains where security is still an emerging concern, such as cars and industrial manufacturing.

To prevent injection attacks during software development, developers need to define the languages of all documents used for communication in a system. Not many developers will take on this complex task unless the language definition process is as simple as writing a string to a file. Therefore, approaches that ease the language definition and (un)parsing during development will help to prevent injection vulnerabilities as well.

## IV. Defining correct Unparser and Encoder

Following the approach of producing documents by unparsing, let us consider a context-free grammar definition of a language $L$ and an AST $x$ of a document written in $L$. Program logic developers use the AST as an interface to the document written in $L$, so they store input data into the data tokens of $x$ at the places where this input data should be placed within the output document. Hence maliciously crafted input data aiming at injecting control tokens into $L$ is stored into these data tokens as well.

By utilizing the regular expression that defines the data token within the grammar, an unparser can validate if the data token complies with the definition of the token. To prevent an injection vulnerability, the unparser can simply stop unparsing this document. Although this approach would prevent injections for tokens where the regular expression is formulated sharp enough, it ultimately prevents control tokens from being used within legal input data, e.g. books published by O'Reilly won't work in a SQL query.

```
1  package de.se_rwth.format;
2
3  grammar Container {
4    options {
5      nostring nomlcomments noslcomments
6      noident lexer lookahead = 4
7    }
8
9    Body = LCURLY Element* RCURLY;
10   Element = "tags"
11        LCURLY TagsToken RCURLY;
12   token LCURLY = "{"; token RCURLY = "}";
13
14   encodeTable TagsToken = {
15     "{" -> "ģ", "}" -> "ĥ",
16     "&" -> "8", " " -> " "
17   };
18
19   subparser token TagsToken =
20     (~('{' | '}' | ' '))+
21   ;
22 }
```
Listing 1. MontiCore Grammar example with integrated encoder and sub parser definition

To bypass this limitation, escape sequences are used to encode control tokens within data tokens. As mentioned before, escape sequences are context-specific, and a context corresponds to a token in the language's grammar. Hence, in our approach, we allow language developers to add an encoding table for every token defined in the grammar. This table defines the context-specific encoding for that token using encoding rules that translate a control token to an escape sequence within the data token. As shown by the encoding table for the token `TagsToken` in the example grammar in listing 1, we integrate the encoding table definition into the grammar.

Defining encoding rules is a crucial step during language definition because control tokens that are missing in the encoding table or whose encoding results in another control token lead to injection vulnerabilities in the unparser that is derived from the encoding table. To make this critical task easier, a default encoding table that encodes all control tokens not allowed in a data token is much appreciated. This would prevent errors in the encoding definition.

While the unparser context-sensitively encodes data tokens when transforming an AST into a document, the parser needs to decode data tokens when reading a document to enable a correct (un)parser round-trip.

```
1  package de.se_rwth.format;
2
3  grammar Tag {
4    options {
5      nostring nomlcomments
6      noslcomments noident
7    }
8
9    Tags = Tag (COMMA Tag)*;
10
11   Tag = LT TEXT GT;
12
13   token LT = "<";
14   token GT = ">";
15   token COMMA = ",";
16   token TEXT =
17     (
18       ~('<' | '>' | '\\' | ',')
19     |
20       (
21         '\\'
22         ('<' | '>' | '\\' | ',')
23       )
24     )+
25   ;
26
27   encodeTable TEXT = {
28     "\\" -> "\\\\",
29     "," -> "\\,",
30     ">" -> "\\>",
31     "<" -> "\\<"
32   };
33 }
```
Listing 2. MontiCore Grammar example with integrated encoder definition

### A. Language Composition

Documents that contain different kinds of information, e.g. structural and behavioral information, are commonly written in a composed language, e.g. a composition of HTML and JavaScript. One way to compose a super-language $A$ and sub-language $B$ is by replacing a nonterminal in the grammar of $A$ with the start symbol of $B$'s grammar. In this way, a parser constructed from the composed grammar accepts documents of the composed language.

When unparsing a composed language, the order in which encoders defined in both languages are applied is critical to achieving a correct (un)parse round-trip.

The encoder application has to start at the most nested language $B$, which is unparsed to a string and placed into the AST data token of the outer language $A$ where $B$ is embedded. Then the unparser of $A$ applies the encoding defined by the tables in $A$ to its tokens, so that the data token that contains $B$ is encoded again, if an encoding table is defined for it. This encoding process is applied for each language - from the most deeply nested to the outmost. By using this application order of encoders, a token from the sub-language is encoded as defined by its encoding table, and afterwards the whole string resulting from the unparsed sub-language is encoded to fit into the data token of the super-language. So in essence the sub-language token is encoded first by its encoder and then by the encoder of the data token of the super-language, where the sub-language is imbedded. Listing 1 and 2 show an example of a composed language, where the keyword `subparser` is used to mark a token as containing a sub-language.

Of course the order in which encoders are applied needs to correspond to the order in which decoders are applied when parsing the composed language. Here, it is important to first parse the super-language, then decode its tokens, and after decoding parse the sub-language, as the decoding changes escape sequences back to control tokens of the super-language, which might influence the parsing of the inner language. This way control tokens of the super-language may be used in an encoded representation in the sub-language so that the parser of the super-language is not influenced by its sub-language. This allows, for example, to easily embed a language into its own data token after having correctly defined the encoding table of that token.

### B. Reducing Language Features

Up to now we focused on injections that modify the structure of the document, also known as 'injecting up' [30]. Now let us take a look on 'injecting down', where control tokens are injected into a context to modify the meaning of the statement without breaking out of the context. For example when developers intend to let users create a part of the output document, e.g. by allowing HTML in forums or wikis. This user input needs to be parsed and validated to remove tokens which a user should not be able to inject into the output document. Just like a firewall can be configured to filter out unwanted network traffic, a filter for unwanted tokens is needed.

There are at least two options to solve this problem: First, the grammar of the output document can be reduced to allow only a subset of the original language. This is appropriate if there is a really nasty feature of the output language that neither the user or developer should use. This kind of reduction does not help in many cases as most developers will need to use all language features at some point. Hence, the grammar needs to be rewritten, such that the reduced language is only used for some tokens, e.g. HTML-tags named in a specific way. This can be achieved by changing grammar productions or by adding an encoding table and a sub-language which encodes the tokens not allowed in the user input.

As rewriting the output document's grammar based on properties of tokens is not a very elegant or easy solution, an alternative is to define a grammar for the reduced language and use it to recognize and validate the input during parsing independently of the unparsing of the output document. This validated input is then copied from the input AST into the output document's AST. In this approach, we use the reduced input language to prevent 'injecting down' and the non-reduced output language to prevent 'injecting up'.

Here, we have a usability trade-off: Reducing the output grammar is a safe way to embed input data into an output document, as all inputs, no matter where they come from, are correctly unparsed into the output document. However, it is not as handy as using the reduced grammar to filter the input. Therefore, for systems where the input is directly used within the output without modification, the latter approach appears to be suitable.

In contrast, for systems with a complex processing logic, which changes the input or enriches it using data from multiple sources, a reduced output grammar is inevitable. This is because the processing logic inserts data that the unparser assumes to be valid, but in fact may contain forbidden parts of the output language. This issue cannot be fixed by validating all inputs when they enter the system, as the validation needs to consider the context where the input is used in the output document.

### C. Implementation

To show the feasibility of our approach, we implemented it within the MontiCore framework. The framework generates parsers, unparsers, encoders, and decoders from the context-free MontiCore grammar, which defines tokens that contain a sub-language, e.g. line 19 of listing 1, and encoding tables, e.g. in line 27-32 of listing 2 for the `TEXT` token.

Although the compact representation of an encoding table within the MontiCore grammar is useful when developing a new format, it is not when encoder libraries already exist or the encoding can be written more compact in program code. In these cases, the implementation allows for the replacement of generated encoders for specific tokens with encoding functions defined in Java code. To limit the effect of errors in the encoding table definition and encoder implementation from a library, an encoded data token is validated using the regular expression that defines the token.

This check may prevent an injection, if the token definition is strictly defined and does not allow control tokens.

An issue when developers use this approach is that creating an AST using program code is quite tedious.

So to improve the applicability of the approach we use templates like the one in listing 3 to define a basic structure of the output document's AST. The template needs to be a valid document of the language, as we use the generated parser to create the AST from it. Variables like `#name#` in the template end up in data tokens within the AST and are replaced with (input) data provided by the developer. To add more control and data tokens to the AST, the developer can alter the AST before unparsing it.

## V. Evaluation

We evaluate our solution by taking an example of HTML and JavaScript, as XSS is a prominent problem, and there are tools for testing applications for XSS.

For both languages, we developed a grammar that we composed to define the combined language.

In HTML there are different contexts where JavaScript is embedded and encoded in different ways, which makes the definition of the language more complex. The symbols `<`, `>`, and `&` need to be encoded in every HTML-tag's body except a script-tag's body. As a result, in the HTML grammar two different contexts and hence tokens have to be defined for a tag's body depending on the tag's name: one for a script tag body and another one for all other HTML-tag bodies.

```
1  <?xml version="1.0" encoding="UTF-8"?>
2  <!DOCTYPE html>
3  <html xmlns="http://www.w3.org/1999/xhtml">
4    <head> <title>Example Page</title> </head>
5    <body>
6      <form method="GET" action="#actionURL#">
7        <label for="input_name"> Name: </label>
8        <input type="text" id="input_name"
9          name="name" value="#name#" />
10       <input type="submit" value="Register"/>
11     </form>
12     <div>
13       <p>#name#</p>
14       <button
15         onclick="alert("#name#");"
16       >Test1</button>
17       <button onclick="alert('#name#');"
18       >Test2</button>
19       <input type="text" name="input"
20         value="#name#" />
21       <script>
22         var name ="#name#" + '#name#';
23       </script>
24     </div>
25   </body>
26 </html>
```

Listing 3. HTML/JavaScript template with markers *actionURL* and *name*

To make things worse, HTML is also inconsistent with contexts where HTML and JavaScript are allowed. Within a tag's attribute, the symbols `<`, `>` and `&` as well as quotation marks must be encoded. However, in attributes containing JavaScript, like the onclick-attribute, those characters have to be encoded like in every other attribute. This results in `alert('"');` opening a pop up message `"`, if it is used in a script tag's body and a message with a quotation mark when used in an onclick-attribute. Furthermore the comparison (`5 < 6`) works in an onclick-attribute but not in a script-tag's body. We successfully implemented all these special cases using encoding tables.

Utilizing the generated parser, decoder, encoder, and unparser, we developed a web application that provides an HTML form with input fields. To produce an HTML output document that includes the input data at different contexts, the template in listing 3 is used to set up the AST, and data tokens marked with `#...#` are replaced by input data.

This setup enables us to use manual and automated penetration testing (the standard approach for finding input validation vulnerabilities in web applications) to test our generated encoder and decoder. We use the Zed Attack Proxy (ZAP) [31] as an interception proxy to test the web application for reflected cross-site scripting vulnerabilities. As a first step, we used the "Active Scan" to automatically find vulnerabilities by sending XSS strings to the web application. Therefore, we configured the "Cross Site Scripting (Refelected)", "CRLF injection", and "Parameter tampering" option with threshold for notifications to "Low" and set the strength to "Insane". In this configuration, ZAP sent 100 attack strings and raised two alarms. Both alarms are raised for the input `;alert(1)`, and the output of the test application is shown in listing 4

```
1  <button onclick=
2      "alert(";alert(1)");">
3    Test
4  </button>
5
6  <button onclick=
7      "alert(';alert(1)');">
8    Test
9  </button>
```

Listing 4. ZAP alerts found with active scan

These are false positives, as both values are correctly encoded JavaScript string literals enclosed in an HTML attribute.

In a second step, we used the XSS attack strings from FuzzDB, which are integrated to ZAP to manually test the application. We grouped the attack strings for similar attacks and chose several strings from each group resulting in a subset of eleven strings. These strings are submitted to all output contexts marked in listing 3 and checked for correct presentation in a browser. All tested strings from FuzzDB were encoded correctly. Therefore, the browser displayed them and did not execute them as JavaScript.

## VI. Conclusion and Future Work

Software that consumes input and uses this input to produce documents written in a formal language faces the problem of preventing input injection in the document. Well-known and widespread vulnerabilities that arise when input is injected are typically named after the language of the document, e.g. SQL injection, or cross-site scripting (XSS) in case of HTML and JavaScript. Such naming masks the common nature of these vulnerabilities.

To prevent input injection, the input data has to be encoded so that it does not contain control tokens of the document. However in many cases documents have multiple contexts where input data has to be encoded differently. Hence, data encoding cannot be done during input parsing but has to be applied during the document creation depending on the context within the output document.

In this paper we worked on the problem of correct unparsing and encoding maliciously crafted user data into a document whose language is defined by a context free grammar. In addition, we examined how parts of the output document can be defined by input data using only a subset of the output document's language features.

We presented an approach to define encoders and decoders along with the language's grammar and used the MontiCore framework to implement a generator that derives context-sensitive encoders and decoders from that definition to correctly encode maliciously crafted data during document creation. For the safe definitions of parts of the document using input data, we discussed the approaches of reducing the output language and validating the input along with their advantages and disadvantages. We evaluated the generated encoder and decoder by taking the example of HTML and JavaScript using the penetration testing tools ZAP and FuzzDB and found no context where cross-site scripting was possible.

The presented approach allows developers to prevent input-based attacks such as XSS on all context-free languages they are using as an output, once the language and its encoding rules are defined.

Clearly we assume an idealized world here, where unparsers and parsers are generated from the same grammar. Considering the differences in parser implementations adds another dimension to the problem of correct unparsing and encoding. Following the presented approach, we plan to evaluate the implemented framework for various custom formats defined by developers to study the use of the framework in software development projects. In addition we plan to support default encoding tables and search for ways to improve the auditing of input data processing.

## Acknowledgment

The authors wish to thank Sergey Bratus, and the anonymous reviewers, whose detailed comments greatly improved the presentation of this work, as well as Benedikt Westermann for finding an XSS vulnerability in an early version of the encoding table for the HTML grammar.

This work has been funded by the Excellence Initiative of the German federal and state governments within the HumTec Project House at RWTH Aachen University.